\documentclass[12pt]{article}
 \usepackage[top=2.0truecm, bottom=2.2truecm, left=1.6truecm, right=1.6truecm]{geometry} 

\def \etal      {\hbox{et al.} }

\def \sims      {\sim \!}

\def\\{\hfil\break}
\def\spose#1{\hbox to 0pt{#1\hss}}
\def\lta{\mathrel{\spose{\lower 3pt\hbox{$\mathchar"218$}}
     \raise 2.0pt\hbox{$\mathchar"13C$}}}
\def\gta{\mathrel{\spose{\lower 3pt\hbox{$\mathchar"218$}}
     \raise 2.0pt\hbox{$\mathchar"13E$}}}

\newcount\itemno
\def \ino         { \the\itemno\global\advance\itemno by 1 }
\def \bref{\par \noindent \hangindent=1.5 truecm \hangafter=1}

\begin{document}


\begin{center}
{\large \bf Dark Energy Studies: Challenges to Computational Cosmology}\\
\smallskip
{\normalsize James Annis, Francisco J. Castander, August E. Evrard, Joshua A. Frieman, \\ 
Enrique Gaztanaga, Bhuvnesh Jain, Andrey V. Kravtsov, Ofer Lahav, Huan Lin, Joseph Mohr, \\ 
Paul M. Ricker, Albert Stebbins, Risa H. Wechsler, David H. Weinberg , Jochen Weller\\
{\sl theory members of the DES collaboration}}
\end{center}

 \abstract
The ability to test the nature of dark mass-energy components in the universe through large-scale structure studies hinges on accurate predictions of sky survey expectations within a given world model.  Numerical simulations predict key survey signatures with varying degrees of confidence, limited mainly by the complex astrophysics of galaxy formation.   As surveys grow in size and scale, systematic uncertainties in theoretical modeling can become dominant.   Dark energy studies will challenge the  computational cosmology community to critically assess current techniques, develop new approaches to maximize accuracy, and establish new tools and practices to efficiently employ globally networked computing resources.  

\section{Introduction}

Ongoing and planned observational surveys, such as the Dark Energy Survey (DES)$^{\ino}$, are  providing increasingly rich information on the spatial distributions and internal properties of large numbers of galaxies, clusters of galaxies, and supernovae.  These astrophysical systems reside in a cosmic web of large-scale structure that evolved by gravitational amplification of an initially small-amplitude Gaussian random density field.  The DES plans to investigate the dark sector through the evolution of the Hubble parameter $H(z)$ and linear growth factor $D(z)$ from four independent channels:  i) the evolution and clustering properties of rich clusters of galaxies; ii) the redshift evolution of baryonic features in the galaxy power spectrum; iii) weak-lensing tomography derived from measurement of galaxy shear patterns and iv) the luminosity distance--redshift relation of type Ia SNe.  We focus our attention on theoretical issues related to the first three of these tests.  

The power spectrum of fluctuations at recombination is calculated to high accuracy from linear theory$^{\ino}$, so the problem of realizing, through direct simulation, the evolution of a finite patch of a particular world model from this epoch forward is well posed.  To support DES-like observations, one would like to evolve multiple regions of Hubble Length dimension with the principal clustered matter components --- dark matter and multiple phases of baryons (stars and cold gas in galaxies, warm/hot gas surrounding galaxies and in groups/clusters) --- represented by multiple fluids.   Mapping observable signatures of the theoretical solution along the past light-cone of synthetic observers in the computational volume allows `clean' mock surveys to be created, which can further be `dirtied' by real-world observational effects in support of survey data analysis.  

Two fundamental barriers stand in the way of achieving a complete solution to this problem.  One is the 
wide dynamic range of non-linear structures (sub-parsecs to gigaparsecs in length, for example), and the other is the complexity of astrophysical processes controlling the baryonic phases.  Since DES-like  surveys probe only the higher mass portion of the full spectrum of cosmic structures, the first issue is not strongly limiting.  The DES will, however, require understanding galaxy and cluster observable signatures, so uncertainties in the treatment of baryonic physics will play a central role.   In a companion paper$^{\ino}$, we outline theoretical uncertainties associated with the large-scale structure channels DES will use to test dark energy.  Here, we offer a critique of the computational methods that  provide theoretical support for DES and similar surveys. 

\section{Challenges for Computational Cosmology } 

Given the wide scale and scope of large-scale structure, a number of approaches have evolved to address restricted elements of the full problem.  Since cold dark matter dominates the matter density, N-body methods that follow collisionless clustering have played an important role in defining the overall evolution of the cosmic web and the structure of collapsed halos formed within it.   Combined N-body and gas dynamics techniques explore gravitationally coupled evolution of baryons and dark matter.   Knowledge gained from these `direct' approaches led to the development of  `semi-analytic' methods that efficiently explore scenarios for baryon-phase evolution.   The overall  challenge to computational cosmology in the dark energy era is to understand how to harness and push forward these different methods so as to maximize science return from sky survey data. 

A `halo model' description of the large-scale density field ties together these approaches.  The model posits that all matter in the late universe is contained in a spectrum of gravitationally bound objects (halos), each characterized by a mass $M$ defined (typically) by a critical overdensity condition around a local filtered density peak.  The space density, large-scale clustering bias (relative to all matter), and internal structure, such as density and temperature profiles, are basic model ingredients.  For galaxy studies, the halo occupation distribution (`HOD') defines the likelihood $p(N_{gal}|M,z)$ that $N_{gal}$ galaxies of a certain type are associated with the halo.  For some applications, it may be important to consider HOD dependence on local environment. 

\vspace{-0.2truecm}
\subsection{Collisionless N-body modeling} 

Understanding the growth of density perturbations into the mildly and strongly non-linear regimes is a critical component for  weak lensing tomography and galaxy cluster studies, respectively.   Much progress has been made in this area using N-body simulations, as Moore's law has enabled progressively larger computations, up to $10^{10}$ particles today$^{\ino}$.   Parallel computing has led to production environments where $512^3$ particle runs can be realized on an almost daily basis.   

By creating large statistical samples of dark matter halos and by probing the internal structure of some halos in great detail, large-$N$ simulations have validated and characterized important raw ingredients of the halo model:  i) the space density is calibrated to $\sims 10\%$  accuracy in terms of a similarity variable $\sigma(M,z)$, with $\sigma$ the {\sl rms} level of density fluctuations filtered on mass scale $M$;  ii) the large-scale clustering bias of halos is calibrated to similar accuracy;  iii) except for rare cases of major mergers in progress, the interior of halos are hydrostatic with an internal density profile that depends primarily on mass, and secondarily on individual accretion history and iv) the structural similarity of halos is reflected in a tight virial scaling between mass and velocity dispersion.   The study of sub-halos, locally bound structures accreted within larger halos but not fully tidally disrupted, is a rapidly developing area with important application to optical studies of galaxy clusters.   Since they serve as a foundation for more complex treatments, these ingredients of the halo model deserve more careful study and more precise calibration. 

The weak lensing shear signal on arcminute and larger scales is generated by weakly non-linear matter fluctuations on relatively large spatial scales, making it relatively insensitive to small-scale baryon physics.  Although the Hamilton-Peacock characterization$^{\ino}$ of the non-linear evolution of the power spectrum has been useful, it will need refinement to achieve the anticipated accuracy of DES power spectrum measurements on dark energy.  The evolution of higher density moments, particularly the bi-spectrum, is less well understood than the second moment.  A suite of multi-scale N-body simulations covering a modest grid of cosmological models is needed to address these problems.  New approaches to generating initial conditions and combining multiple realizations of finite volumes$^{\ino}$ should be employed in an effort to push systematic uncertainties on relevant spatial scales to percent levels and below.  

Although a relatively mature enterprise, N-body modeling of dark matter clustering faces fundamental challenges to improve the absolute level of precision in current techniques and to better understand the dynamical mechanisms associated with non-linear structure evolution.   Code comparison projects$^{\ino}$  should be more aggressively pursued and the sensitivity of key non-linear statistics to code control parameters deserves more careful systematic study.  A return to testing methods on the self-similar clustering problem$^{\ino}$ is likely to provide valuable insights, and deeper connections to analytic approaches,  such as extended perturbation theory and equilibrium stellar dynamics, should be encouraged.  Highly  accurate dark matter evolution is only a first step, however, as it ignores the $17\%$  matter component of the universe that is directly observable.   

\vspace{-0.2truecm}
\subsection{The baryon phase and galaxy/cluster observables }  

The dark energy tests planned by DES require modeling the astrophysical behavior of different baryonic phases.   Acoustic oscillations in the galaxy power spectrum must be linked to features in the matter power spectrum, requiring accurate tests of the constancy of galaxy bias on large scales.  For clusters, selection by Sunyaev-Zel'dovich or X-ray signatures depends on the hot gas phase properties while optical selection is dependent on star formation and interstellar medium phase evolution within galaxies.   Several distinct, but related, approaches have emerged to address this complex modeling requirement, all involving tunable model parameters that must, to some degree, be determined empirically.

`Direct' computational approaches couple a three-dimensional gas dynamics solver to an N-body algorithm.  A dozen, nearly-independent codes, in both Lagrangian and Eulerian flavors, now exist to perform this task.  All methods follow entropy generation in gas from shocks, and most allow radiative entropy loss assuming local thermodynamic equilibrium in plasma that may optionally be metal enriched.  Methods diverge in their treatment of interstellar medium processes: cold-hot gas phase interactions, star formation rate prescriptions, return of mass and energy from star forming regions, supermassive black hole (SMBH) formation, and attendant SMBH feedback.   A valuable comparison study$^{7}$ revealed  agreement at the $\sim 10\%$ level among a dozen codes for the solution of the internal structure of a single halo evolved without cooling and star formation.  

In massive halos, the hosts to rich clusters where only a small fraction of baryons condense into galaxies,  gas dynamic models have had good success in modeling the behavior of the hot intracluster medium (ICM), particularly its structural regularity.  The ICM mass, a quantity that is essentially independent of temperature when derived from low energy X-ray surface brightness maps, is observed to behave as a power-law of X-ray temperature with only $14\%$ intrinsic scatter$^{\ino}$.   Simulations have been instrumental in showing that this tight scaling relation results from a combination of factors:  i) approach to virial equilibrium is rapid, so temperature and total halo mass are strongly correlated; ii) the ICM mass fraction varies by $\lta 10\%$ at a given mass, meaning variations in local baryon fraction and galaxy formation efficiency at fixed mass are small and iii) the ICM plasma outside the core is not strongly multi-phase, making the spectral temperature a good indicator of the host halo's gravitational potential.   

Although the sensitivity of the hot ICM to galaxy formation and galactic nuclear feedback is under active investigation$^{\ino}$, predictions from direct simulations for observable scaling relations are not likely to converge fast enough to be useful for DES cluster analysis.  Instead, such simulations can offer immediate support by constraining the expected forms of the likelihoods for how properties like X-ray temperature, intrinsic galaxy richness or total  SZ decrement should scale with mass and epoch.  Power-law distributions with log-normal scatter are supported by current simulations, but more study is needed to understand potential deviations from power-law behavior, the origins of the variance, and the covariance of independent observable measures.   For a sufficiently rich data set, the parameters describing the observable-mass relations (log-mean slope and intercept, dispersion, and drifts in these with scale factor) can be folded into the analysis as nuisance parameters and solved for directly using differential survey counts and clustering properties.  Fisher matrix analyses$^{\ino}$ offer reasons to be optimistic  about this self-calibrating approach, but studies based on mock sky survey data, where projection effects can lead to more complex scatter in observable scaling relations, remain to be done.  

Because baryon oscillation and detectable weak lensing effects are tied to scales significantly larger than individual galaxies, these dark energy probes should be relatively insensitive to uncertainties in galaxy formation physics.  Direct simulations are inefficient at testing this assumption, since large volumes need to be modeled at very high spatial resolution, and many realizations covering a range of control parameters need to be done.  Instead, hybrid approaches are favored that combine large N-body models, or equivalent halo model realizations, with `semi-analytic' or empirically-derived HOD galaxy assignment schemes.  

Semi-analytic methods use simplifying assumptions, calibrated by direct simulation, to reduce the problem of galaxy formation to a set of coupled ordinary differential equations describing the flow of mass and energy among different baryonic components.  Their flexibility and computational efficiency allow large regions of the controlling parameter space to be explored, a significant advantage compared to direct methods.  Semi-analytic models reproduce much, but not yet all, of the full statistical richness of the galaxy populations in local (SDSS and 2dF) catalogs.   There are questions regarding uniqueness of the prescriptions for the various flows and it is not yet clear how degenerate are the existing control parameters (which number in the several tens at the least).  In addition, predicting optical and near infrared signatures entails modeling galaxy stellar populations and dust content, introducing further complexity and uncertainty.  An important issue for dark energy is to understand the sensitivity of the galaxy power spectrum to different treatments of galaxy formation, so alternatives to semi-analytic and direct modeling deserve consideration.  Methods that use empirically-trained HOD or similar statistical schemes to relate galaxies and dark matter offer a potentially powerful and complementary means to explore systematic uncertainties in galaxy/cluster-based dark energy studies.  

\vspace{-0.2truecm}
\subsection{Mock surveys as theory testbeds}

Correct interpretation of sky survey data requires detailed understanding of survey selection and signal contamination issues.  Mock surveys provide a testbed for assessing the purity and completeness of well-defined samples and for exploring the influence of non-trivial contamination from projection.  
By offering a highly realistic input signal, to which observational effects like survey masks and instrumental noise can be added, mock surveys can also provide the important service of end-to-end testing of data processing and analysis pipelines.   

Dark energy tests from weak lensing and baryon features in the galaxy power spectrum require distance-selected galaxy samples.  For DES, distance will be based on photometric redshifts derived from broad-band colors.  The accuracy of photometric redshifts is under ongoing empirical investigation, but the distribution of errors is likely to be non-Gaussian in at least some regions of color space.  Mock surveys can address the impact of complex photo-z errors on derived measurements, and offer insights into selection procedures that maximize signal to noise.  

For cluster detection, projected contamination is a potentially serious concern.  At sub-mm wavelengths, simulated sky maps from direct simulations indicate that planned experiments will be highly complete above a halo mass roughly one-fifth that  of Coma.   Optical cluster detection is more susceptible to contamination, and studies aimed at calibrating purity and completeness of particular methods remain under development.   The relation between galaxy color and local density is a crucial element.  Since this relation is not particularly well reproduced by direct or semi-analytic methods, parametric approaches  that consider the likelihood of a galaxy of a particular color inhabiting some location (defined by dark matter properties like local mass density or constituent halo mass) may be more powerful.  

An exercise that has not yet been done is to baseline the current level of uncertainties by comparing sky survey expectations generated by multiple methods --- N-body with semi-analytic galaxy assignment, N-body with empirical galaxy assignment, and direct gas dynamic simulation --- within a fixed world model.  For clusters, the question of how best to combine optical, sub-mm and X-ray observations to provide optimal constraints on dark energy needs more careful study.  

\section{Opportunities from Emerging Technologies} 

Progress in computational cosmology accelerated when groups of researchers pooled intellectual resources and formed consortia, such as GC$^3$ and the N-body Shop in the US and the Virgo Consortium in Europe.  Consortia offer advantages, in the form of shared computational frameworks and systems for its members, within which large-scale efforts can be realized more efficiently.  In the US, the `grand challenge' era of consortia building was driven by investments in supercomputer resources by NSF and NASA in the mid-1990's.  The emergence of relatively low-cost parallel clusters coincided with the transformation of US federal support into a partnership model, in which universities were expected to provide substantial investment in computational infrastructure.  Given the many distractions caused by the `dot-com' bubble, most universities were slow to respond.  As a result, the landscape today is highly fractured.  While serious computing power exists at select locations (in the US, primarily national laboratories and a few, highly subscribed supercomputer centers), most research groups are hard pressed to put together and maintain a compute cluster with a few tens of processors.  

In recent years, strong support for bandwidth-oriented development has created a new atmosphere of opportunity for computational research.  Within the nascent world of Grid computing lie a number of rational elements to improve research productivity, such as  intelligent networked resource brokers and schedulers, dedicated ``quality of service'' bandwidth, and single sign-on authorization.   The computational cosmology community should take a more active role in establishing the `middle-ware' elements needed to enhance productivity within collaborations and the sharing of information within and among them.  

The building of Theory Virtual Observatory archives, with libraries of open source codes, raw resources (particle/cell-level files), and processed resources (halo catalogs and histories, sky survey catalogs, sky maps) needs to get underway in earnest.  Dark energy science demands theory support at the highest level and of the highest quality.  The time has come to leverage the networked resources of the information age and build a cyberinfrastructure framework that will bring theory and observation closer together, enhancing dark energy and astrophysical science studies in the process. 

\bigskip
\noindent
{\bf References}

{\small 
\itemno=1
\bref $^{\ino}$ 
See {\tt http://www.darkenergysurvey.org}
\bref $^{\ino}$ 
Seljak, U., {Sugiyama}, N., {White}, M., {Zaldarriaga}, M. 2003, PhRvD, 3507
\bref $^{\ino}$ 
J. Annis \etal 2005, Theory White Paper to the Dark Energy Task Force
\bref $^{\ino}$ 
Springel, V. \etal 2005, Nature, 435, 629
\bref $^{\ino}$ 
Sirko, E. 2005, ApJ, submitted, ({\tt astro-ph/0503106})
\bref $^{\ino}$ 
 {{Hamilton}, A.J.S.,  {Kumar}, P., {Lu}, E. and {Matthews}, A.} 1991, ApJ, 374, L1;  {Peacock}, J.A. and {Dodds}, S.J. 1996, MNRAS, 280, L19;  Smith, R.E. \etal 2003, MNRAS, 341, 1311.
\bref $^{\ino}$ 
Frenk, C.S. \etal 1999, ApJ, 525, 554
\bref $^{\ino}$ 
{{Efstathiou}, G., {Frenk}, C.S., {White}, S.D.M. and {Davis}, M.} 1988, MNRAS, 235, 715
\bref $^{\ino}$ 
 {{Mohr}, J.J.,  {Mathiesen}, B. and {Evrard}, A.E.} 1999, ApJ, 517, 627
\bref $^{\ino}$ 
{{Kravtsov}, A.V., {Nagai}, D. and {Vikhlinin}, A.A.}, 2005 ApJ, 625, 588;  Sijacki, D and Springel, V. 2005, astro-ph/0509506;  Borgani, S. \etal 2004, MNRAS, 348, 1078
\bref $^{\ino}$ 
Lima, M and Hu, W. 2005, PhRvD, 72, 3006;  Lima, M and Hu, W. 2004, PhRvD, 70,  3504
}

\end{document}